\documentclass[conference]{IEEEtran}
\usepackage{array}
\usepackage{booktabs}
\usepackage{listings}

\usepackage[shortlabels, inline]{enumitem}
\usepackage{graphicx}
\usepackage{xcolor}
\usepackage{fancyvrb}
\usepackage{tabu}
\usepackage{tabularx}
\usepackage{array}
\newcolumntype{L}[1]{>{\hsize=#1\hsize\raggedright\arraybackslash}X}%
\newcolumntype{C}[1]{>{\hsize=#1\hsize\centering\arraybackslash}X}%
\newcolumntype{R}[1]{>{\hsize=#1\hsize\raggedleft\arraybackslash}X}%
\usepackage{booktabs}
\usepackage{todonotes}
\usepackage{marginnote}

\usepackage{amsmath}
\usepackage[numbers]{natbib}
\usepackage{flushend}
\usepackage[hidelinks]{hyperref}
\usepackage{cleveref}

\usepackage{ifthen}
\usepackage{standalone}
\usepackage{tikz}
\usetikzlibrary{shapes,backgrounds}
\usetikzlibrary{decorations.text}
\usepackage{pgfplots}
\pgfplotsset{compat=1.12}
\crefname{table}{tab.}{tab.}
\crefname{figure}{fig.}{fig.}

\usepackage{float}
\begin{document}
\title{Improving Function Coverage with Munch: A Hybrid Fuzzing and Directed Symbolic Execution Approach}

\author{\IEEEauthorblockN{Saahil Ognawala, Thomas Hutzelmann, Eirini Psallida, Alexander Pretschner}
    \IEEEauthorblockA{\textit{Technical University of Munich, Germany} \\
        \{saahil.ognawala, t.hutzelmann, eirini.psallida, alexander.pretschner\}@tum.de}
}

\maketitle
\begin{abstract}\label{sec:abstract}
Fuzzing and symbolic execution are popular techniques for finding vulnerabilities and generating test-cases for programs. Fuzzing, a blackbox method that mutates seed input values, is generally incapable of generating diverse inputs that exercise all paths in the program. Due to the path-explosion problem and dependence on SMT solvers, symbolic execution may also not achieve high path coverage. A hybrid technique involving fuzzing and symbolic execution may achieve \emph{better function coverage} than fuzzing or symbolic execution alone. In this paper, we present Munch, an open-source framework implementing two hybrid techniques based on fuzzing and symbolic execution. We empirically show using \emph{nine large open-source programs} that overall, Munch achieves higher (in-depth) function coverage than symbolic execution or fuzzing alone. Using metrics based on total analyses time and number of queries issued to the SMT solver, we also show that Munch is more efficient at achieving better function coverage.
\end{abstract}

\definecolor{orange}{rgb}{1,0.5,0}



\section{Introduction}\label{sec:introduction}
Fuzzing and symbolic execution often do not achieve high coverage, not only at the source code, binary, or any intermediate code levels but also at the \emph{component level}. Components of software are its basic building blocks, such as classes in object oriented programming or functions in functional programming. Component coverage, thus, may be defined as the number of components in a program that are executed \emph{at least once} during a software testing cycle. Low component coverage can be observed particularly frequently for components that can only be reached via many other components in a chain of components \cite{stephens2016driller,banescu2016code}. It is also true for scenarios where components may be reused across programs, e.g., in the form of open APIs \cite{christakis2015ic}, or when they are developed by different teams. Low component coverage intuitively increases the probability that vulnerabilities in uncovered components lying deep inside a program remain undiscovered.  

Automated methods for discovering such vulnerabilities can be classified into two categories. \emph{Blackbox} testing methods are oblivious to the inner workings of the system under test (SUT). \emph{Whitebox} methods, on the other hand, take into account the programs' underlying structure, via source code, software model or any other intermediate representation, to derive efficient testing strategies. Most testing methods in practice can be placed on a spectrum between blackbox and whitebox (also called \emph{greybox}), depending on the available information about the SUTs. \emph{Fuzzing}, or fuzz testing \cite{sutton2007fuzzing}, is an example of a (mostly) blackbox testing methodology. Fuzzing works with and modifies -- often user-supplied -- test-cases to discover new, previously unseen, functionality in a software. The modification (mutation) of test-cases depends on heuristics that are \emph{expected} to help discover new program paths. Due to its oblivion w.r.t.\ SUTs' implementations, fuzzing may miss program paths that are unlikely to be executed with randomly mutated input. Such program paths may even involve branches that are close to one of the entry points of a program, but difficult to enter anyway. 

Symbolic execution (or its practical approaches, e.g., \emph{concolic execution} \cite{sen2005cute} or \emph{whitebox fuzzing} \cite{godefroid2008automated}), a \emph{whitebox} method, is a deterministic method that uses instrumentation to dynamically collect \emph{constraints} representing branching conditions in a program and solves these constraints to generate inputs that execute different paths in a program. However, due to its reliance on SMT solvers and the problem of path explosion \cite{cadar2013symbolic}, symbolic execution may get ``stuck'' in the shallow parts of a program and never reach (and solve) branches that lie deeper. 

In this paper, we introduce a framework that increases \emph{function coverage} at all depths of C-programs by combining fuzzing and concolic execution. We define function coverage in C programs as follows: If a function (other than \texttt{main}) has been called by another (parent) function at least once during all executions of a program, it is said to be \emph{covered}; otherwise, it is said to be \emph{uncovered}. Our technique makes use of KLEE \cite{cadar2008klee} for symbolic execution, and AFL \cite{afl,aflcov} for fuzzing. We have augmented KLEE with a targeted search strategy, based on a shortest path search algorithm (\cref{sec:symex}). We also present evaluation results of our framework on several open-source programs, demonstrating the effectiveness of our method. 

\paragraph*{Problem} Fuzzing and symbolic execution by themselves often do not achieve high function coverage in all depths of a program. \emph{Shallow} coverage means high coverage in parts of a program that are close to the entry point, and low coverage in parts that are more distant from them, or ``deeper'' in the call-graph. Symbolic execution tools tend to achieve shallow coverage in a limited time because the underlying constraint solvers (e.g., SMT solvers) take too long to return. Fuzzing tools, similarly, are unable to cover deeply located functions whose entry points are guarded by branching conditions that are hard to pass. Without proper coverage at all depths, both these techniques may fail to discover non-trivial vulnerabilities. 

\paragraph*{Solution} Our framework, Munch, combines fuzzing and symbolic execution in two ways to increase function coverage:
\begin{enumerate*}
    \item Use fuzzing (with manual seed-inputs) for an initial exploration of the programs and, then, use targeted symbolic execution to explore those functions that were not covered by fuzzing. And,
    \item Use symbolic execution for generating a \emph{functionally diverse} set of test-cases that are, then, used to fuzz the program. 
\end{enumerate*}
We hypothesize that both these hybrid approaches achieve higher function coverage, also of deeply located functions, at a lower cost (time) than fuzzing or symbolic execution alone. 

\paragraph*{Contribution} 
\begin{enumerate*}
\item Responding to a gap in research and tools in this direction, we implement targeted search functionality in KLEE, for reaching (and, then, exploring) arbitrary function entry points. 
\item We implement two hybrid approaches that treat the bottlenecks of SMT solvers by reducing the number of queries issued by the symbolic execution tool. Because of an initial exploratory phase with fuzzer (in one hybrid method), those functions whose entry points could be reached easily are excluded from the state search by symbolic execution engine and, hence, more time can be devoted to harder-to-reach functions. We demonstrate these approaches and evaluate the gain in efficiency on artificially generated and real-world programs. 
\item To the best of our knowledge, we provide the first empirical evaluation of the effectiveness of a hybrid framework on real-world (9 open-source) programs. We show that when running for less time, our approach achieves higher function coverage than both fuzzing and symbolic execution alone, especially deep in the call-graph where both na\"{i}ve methods result in low coverage. The reason is that we can drastically reduce the number of calls to the constraint solver.
\item Finally, to the best of our knowledge, our framework, Munch, is the only open-source hybrid testing framework that can be used out-of-the-box for any C program. 
\end{enumerate*}

The rest of the paper is arranged as follows. On the grounds of artificially crafted programs, we motivate the problem in \cref{sec:motivation}. We describe our techniques and present details of implemented modules of our framework in \cref{sec:methods}. In \cref{sec:evaluation} we discuss our evaluation metrics, results, and their limitations. \Cref{sec:background} discusses related work. We conclude in \cref{sec:conclusion}. \vspace{-0.5em}
\section{Motivation}\label{sec:motivation}
Using artificially generated C programs, we first demonstrate that fuzzing and symbolic execution are ineffective regarding function coverage in large programs. Addressing inadequate function coverage and reliance on SMT solvers, we demonstrate the need for a more efficient way to reach deeper functions. 

\begin{figure}[h]
    \includegraphics[width=\linewidth]{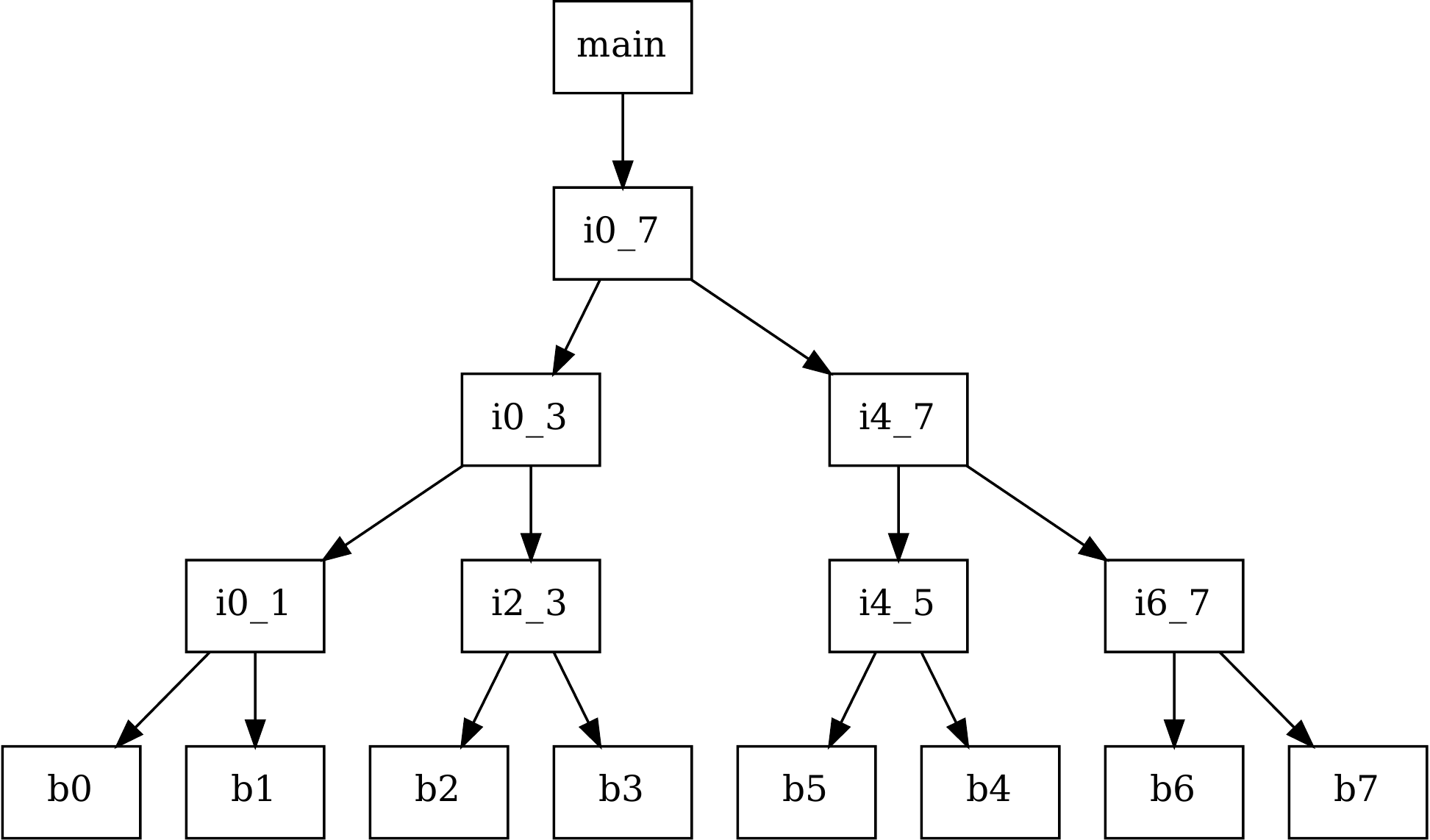}
    \caption{Call-graph of an artificially generated program -- with branching factor of 2 and call-graph depth of 3}
    \label{fig:level3-branching2}
    \vspace{-0.5em}
\end{figure}

We generated C programs with a script that takes as input the desired branching factor, \emph{b}, for all generated functions and the depth of the call-graph of the program, \emph{d}. These programs require users to provide input between $0$ and $b^d-1$. 
The corresponding call-graph of an artificial program with branching factor of 2 and call-graph depth of 3 is shown in \cref{fig:level3-branching2}. For this program, a valid user input lies between 0 and 7 ($2^3-1$). The whole range of integers between upper and lower limit of valid inputs is then progressively split into half, with each half range being the argument to a corresponding function until a function (leaf node in the call-graph) is reached where the input is printed on the screen. For our evaluation, we generated programs for call-graph depths from 1 up to 4, for branching factors 2 up to 4 (total 12 programs). 

Function coverage achieved by fuzzing (AFL) and symbolic execution (KLEE) on all 12 artificially generated programs are shown in \cref{tab:artificial-program-coverage}. AFL does not achieve a desirable function coverage, even for programs that do not contain any complex branching conditions, large input buffers or input-dependent loops. Using KLEE, function coverage is always $100\%$. However, we also need to consider the quantity of SMT solver queries that KLEE issues, for either checking validity of new states, or generating test cases for valid paths \cite{cadar2008klee}. One can see from the $8^{th}$ column of \cref{tab:artificial-program-coverage} that the number of solver queries issued by KLEE is much higher than the number of distinct branches in the corresponding program. 
FS stands for \emph{Fuzzing+Symbolic execution}, which is a simple hybrid technique where, first, the program is fuzzed for a limited amount of time and, then, symbolic execution is used to target those functions which were not covered during fuzzing. We explain this technique and the corresponding improvement achieved w.r.t.\ artificially generated programs in \cref{sec:methods} and \cref{sec:evaluation}.
The number of queries issued by KLEE here is very high and indicates a bottleneck, possibly increasing cost in terms of analysis time.
In fact, as we will see in more detail in \cref{sec:results}, this trend affects real-world programs more severely where complex branching conditions, loops, and external calls are, naturally, more common than in the artificial programs.
\begin{table}[h]
    \centering
    \caption{Function coverage in artificially generated programs}
    \label{tab:artificial-program-coverage}
    \begin{tabularx}{\linewidth}{@{}R{1} C{1} R{.5} R{0.5} R{1} R{1} R{1} R{1} R{1}@{}}
        \toprule
        \rotatebox{90}{Prog. ID} & \rotatebox{90}{Branching} & \rotatebox{90}{Depth} & \rotatebox{90}{Total funcs} & \rotatebox{90}{\parbox{4em}{\% AFL \newline coverage}} & \rotatebox{90}{\parbox{4em}{\% KLEE \newline coverage}} & \rotatebox{90}{\parbox{4em}{\% FS \newline coverage}} & \rotatebox{90}{\parbox{5em}{\# KLEE \newline SMT quer.}} & \rotatebox{90}{\parbox{5em}{\# FS \newline SMT quer.}} \\ \midrule
        P1 & 2             & 1     & 4           & 25     & 100 & 100 & 4648 & 56            \\ \cmidrule(lr){2-2}
        P2 & 2             & 2     & 8           & 25     & 100 & 100 & 4971 & 154          \\ \cmidrule(lr){2-2}
        P3 & 2             & 3     & 16          & 19     & 100   & 100 & 5250 & 63          \\ \cmidrule(lr){2-2}
        P4 & 2             & 4     & 32          & 22     & 100   & 100 & 5783 & 603      \\ \midrule
        P5 & 3             & 1     & 5           & 40     & 100    & 100 & 4790 & 326   \\ \cmidrule(lr){2-2}
        P6 & 3             & 2     & 14          & 36     & 100   & 100 & 5471 & 1169          \\ \cmidrule(lr){2-2}
        P7 & 3             & 3     & 41          & 34     & 100   & 100 & 6390 & 3540         \\ \cmidrule(lr){2-2}
        P8 & 3             & 4     & 122         & 34     & 100  & 100 & 9614 & 6295         \\ \midrule
        P9 & 4             & 1     & 6           & 33     & 100    & 100 & 5248 & 479         \\ \cmidrule(lr){2-2}
        P10 & 4             & 2     & 22          & 32     & 100  & 100 & 5832 & 2716           \\ \cmidrule(lr){2-2}
        P11 & 4             & 3     & 86          & 27     & 100  & 100 & 8327 & 3871          \\ \cmidrule(lr){2-2}
        P12 & 4             & 4     & 342         & 25     & 100 & 100 & 11845 & 6143            \\ \bottomrule
    \end{tabularx}
    \vspace{-1em}
\end{table}

Looking at the performance of fuzzing and symbolic execution we claim that both techniques, by themselves, are inadequate in achieving high functional coverage \emph{efficiently} (i.e.\ overall analysis time). Fuzzing does not generate sufficient diversity in inputs to cover all functions in a program, and symbolic execution spends too much time in SMT solvers. To deal with these limitations of fuzzing and symbolic execution, we now propose a hybrid methodology to achieve high function coverage at all depths of the program in an \emph{effective and efficient manner}. 
\section{Methodology}\label{sec:methods}
Munch\footnote{\small{Munch can be downloaded at \url{https://github.com/tum-i22/munch}}} is an adaptive framework that can operate in \emph{two} hybrid modes. We start this section with a high level description of both modes and, then, give some low level details of the main components of Munch that implement these high level steps. 

\subsection{Modes of Operation}\label{sec:operation-modes} 
\subsubsection{FS Hybrid}\label{sec:fs-mode}
FS stands for \emph{Fuzzing+Symbolic execution}, in that order. In this variant of the hybrid strategy, we fuzz the program for a limited amount of time, compute the function coverage achieved by the fuzzer and, then, use directed (or \emph{targeted}) symbolic execution to reach those functions that were not covered by the fuzzer. When a program is fuzzed for a limited amount of time those functions would be covered, ideally, that are easy to reach (in the program's call-graph) for \emph{most} mutations of the seed-inputs. Using \texttt{sonar-search}, a custom path-search strategy in our symbolic execution engine, as described in \cref{sec:symex}, FS strategy eliminates paths to those functions that are already covered by fuzzing, thereby taking some load off the constraint solver. 

As the reader would have noted, FS hybrid requires the user to provide seed inputs for the first step of fuzzing. Therefore, FS hybrid is particularly useful in those cases where sample inputs for programs are available or can be quickly (manually) generated, especially for file based input.
\subsubsection{SF Hybrid}\label{sec:sf-mode}
For programs where seed inputs may not be easily available or generated, Munch provides a second variant of hybrid greybox fuzzing, called SF hybrid strategy. SF stands for \emph{Symbolic execution+Fuzzing}, in that order. In this variant, we symbolically execute the program under test for a limited amount of time, providing, where needed, symbolic command line arguments, symbolic standard input (\texttt{STDIN}) and symbolic file inputs. With the first round of symbolic execution, hopefully, those functions are covered that are called by diverse inputs. However, due to the bottleneck of constraint solvers, symbolic execution can only achieve shallow coverage. To deal with the shallow coverage of symbolic execution, we use the inputs (standard arguments, \texttt{STDIN} or files) generated by symbolic execution as seed inputs for the fuzzer, as the second step of SF hybrid technique. Due to the diverse, albeit shallow, nature of coverage achieved by symbolic execution, the generated test-cases provide enough diversity in the seed inputs for the fuzzer to trigger different (and, hopefully, deeper) behaviour in the program that is harder to trigger with manually generated seed inputs. 

\subsection{Components of Munch}
To adaptively implement the two modes of operation, as described in \cref{sec:operation-modes}, common cores of fuzzing and symbolic execution are used and the overall operation is managed by our Orchestrator component. We now describe these three components below.  

\subsubsection{AFL for fuzzing}\label{sec:fuzzing}
For fuzzing the programs, Munch employs AFL (American Fuzzy Lop) \cite{afl}, a tool that uses initial test-cases and genetic mutations to generate new test cases. AFL works with target binaries that accept input either directly from \texttt{STDIN} or a file.
AFL receives seed inputs directly from the Orchestrator (\cref{sec:instrumentation}), which is also responsible for adapting the program to convert command line arguments to \texttt{STDIN}, for some special cases.
Our framework, then, fuzzes the program binary for a specific time that we later elaborate in \cref{sec:evaluation}. 

\subsubsection{KLEE for symbolic execution}\label{sec:symex}
Munch adapts and uses KLEE \cite{cadar2008klee} as its symbolic execution engine. For SF mode of operation, that starts with symbolic execution, KLEE is called by the Orchestrator with the path-search strategy of KLEE that is optimized for exploring new unseen paths first \cite{cadar2008klee}. However, for FS mode of operation of Munch, we need a targeted search strategy to arbitrary function entry points.   
Since the original implementation of KLEE does not provide such a strategy, we extended it with a search strategy that prioritizes shortest paths to target functions.

Prototypes, albeit without full implementations or for different languages, for targeted symbolic execution have been proposed in the past \cite{ma2011directed,ognawala2016macke,pretschner2001classical}. Our implementation is inspired by these works, but we modified them so that it takes as input the control-flow-graph and call-graph of a program instead of only original source code, for computing shortest distance. 
Additionally, our implementation uses KLEE's native search heuristic as a \emph{nested searcher} in the following way: after a state has reached the target function, all consecutive states are explored using KLEE's native heuristic. Along the way to the target function, all states that cannot reach the target are terminated immediately, thereby reducing the number of paths to be explored by KLEE. The complete search strategy, with nested KLEE heuristic searcher, is named \emph{\texttt{sonar-search}} in our KLEE fork\footnote{\small{Our KLEE fork can be downloaded at \url{https://github.com/tum-i22/klee22/tree/sonar}}} (because, like sonar waves, we constantly compute distances in the call-graph in a bottom-up fashion and update the searcher). 

\subsubsection{Orchestrator}\label{sec:instrumentation}
The central component that coordinates the working of the fuzzer and symbolic execution engines is the Orchestrator. In addition to this coordination task, Orchestrator is also responsible for generating useful statistics needed for the correct functioning in both modes of operation. Concretely, Orchestrator's functions are as follows
\begin{enumerate}[noitemsep]
    \item Lift the call-graph from a compiled LLVM bitcode of the program\footnote{\small{We only consider the functions that are reachable from \texttt{main}.}}. 
    \item Sort the list of functions in topological order. The reason for topologically sorting the list of all functions is that if \texttt{sonar-search} reaches a frontier node, it is likely that other functions below the level of this frontier node in the call-graph might get covered in the same run of FS hybrid mode. 
    
    A \emph{frontier node} is defined as a function node with a relatively complex guard condition before its entry point and which has a dense sub-(call) graph attached below it. Example of a frontier node is a function that parses an incoming HTTP packet and checks if the packet headers are consistent with the protocol, the \texttt{Content-Length} field conveys true information about the payload etc. Only if the incoming packet header meets the validity criteria defined by an intricate constraint system (guard condition), does the parsing function send the payload down to other processing functions (e.g.\ fetching query results from the backend). 
    
    \item Start Munch by reading a local (to the program under test) configuration file written in JSON format. The mode of operation depends on whether the test-case location contains any text files (FS) or not (SF). 
    \item If the program accepts command line arguments, patch the source-code before compiling to convert command line arguments to \texttt{STDIN}. This is done with a simple wrapper around the \texttt{main} function. 
    \item Initiate fuzzing (FS mode) or symbolic execution (SF mode). 
    \item If operating in the FS mode, read \texttt{afl-cov} \cite{aflcov} coverage information to compute the list of uncovered functions to be used for targeted symbolic execution. 
    \item If operating in the SF mode, read KLEE test-cases and populate the test-case location. 
    \item Initiate symbolic execution (once for every uncovered function) with a user-defined time limit per function (FS mode) or fuzzing (SF mode). 
    \item Finally, compute the function coverage. 
\end{enumerate}
%
%
\section{Evaluation}\label{sec:evaluation}
In this section, we first describe the experiment design to evaluate our methodology w.r.t. function coverage, depth of coverage and the number of calls to the SMT solver, and compare these aspects to pure symbolic execution and fuzzing. Then we discuss our findings in the context of evaluated programs, and the extent to which they may be generalized to other programs.

\subsection{Experiments}\label{sec:experiments}
To demonstrate the effectiveness and efficiency of our approach on real-world programs, we evaluated it on 9 open-source C-programs. \Cref{tab:program-list} lists all evaluated programs with a short description of their respective functionalities. 

\begin{table}[tb]
    \centering
    \caption{List of evaluated programs}
    \label{tab:program-list}
    \begin{tabularx}{\linewidth}{@{}c X@{}} \toprule
        \textbf{Program} & \textbf{Short description}\\ \midrule
        Bc & An interactive mathematical shell\\
        Bzip2 & Data compressor for single files\\
        Diff & Data comparison tool\\
        Flex & Fast lexical analyzer\\
        Grep &  Unix based pattern matching for plain-text files\\
        Jq & JSON file and stream editor\\
        Lz4 & A fast and lossless compression algorithm\\
        Sed & Unix based stream editor\\
        Zopfli & An alternative Gzip compression method\\
        \bottomrule
    \end{tabularx}
\end{table}

For evaluating the effectiveness and efficiency of Munch, we ran experiments with, both, FS and SF modes of operation on all 9 programs. For FS mode, we used a repository of test-cases (manually written by us) to start the first step with AFL. For example, for \texttt{bc}, we created a plaintext file containing the seed inputs, as shown in \cref{list:bc-input-file}.

\begin{lstlisting}[label=list:bc-input-file, caption=Seed input text file for program \texttt{Bc}.]
1+3*7+(2-5)
17/(1+2+3)
8/(1+(2*4+(5+9)))
\end{lstlisting}

For SF mode, we examined the program manuals to gather the minimum requirements w.r.t. lengths of command line arguments, \texttt{STDIN} and/or file inputs and used these lengths on the symbolic input for KLEE. It was essential to use the minimum requirements here because, due to the peculiarities in its modeling of \texttt{libc} environment, the time taken by KLEE to solve symbolic file related constraints increases exponentially with file size, if external \texttt{libc} functions such as \texttt{fgets} or \texttt{fgetc} are called, instead of \texttt{fscanf}. 

Finally, for comparison with pure symbolic execution and fuzzing, we used the same parameters (except timeouts) for KLEE and AFL as we used in the SF and FS respectively. 

The maximum time limits used for analyzing various programs are shown in \cref{tab:time-limits}. We decided to give a bigger timeout value to KLEE and AFL to show that (\cref{sec:results}), even in almost half as much allowed time, Munch is more effective at discovering new functions. Please note that these timeout values were \emph{not} always reached, especially in FS mode, which finished earlier. 
\begin{table}[tb]
    \centering
    \caption{Time limits for all four methods}
    \label{tab:time-limits}
    \vspace{0.2em}
    \begin{tabularx}{\linewidth}{@{}r X@{}}
        \toprule
        Method                                     & Time \\ \midrule
        Fuzzing                                    & 5 hours           \\
        Symbolic execution                         & 5 hours           \\
        FS hybrid & 1 hour (Fuzzing) + \newline 2 mins (\texttt{sonar-search}) per \newline uncovered function \\
        SF hybrid & 1 hour (Symbolic execution) + \newline 1 hours (Fuzzing) \\
        \bottomrule
    \end{tabularx}
\end{table}
\subsection{Results}\label{sec:results}
We now evaluate the effectiveness and efficiency of Munch.

\subsubsection{Effectiveness} 
\begin{figure}[h]
    \centering
    \includegraphics[width=\linewidth]{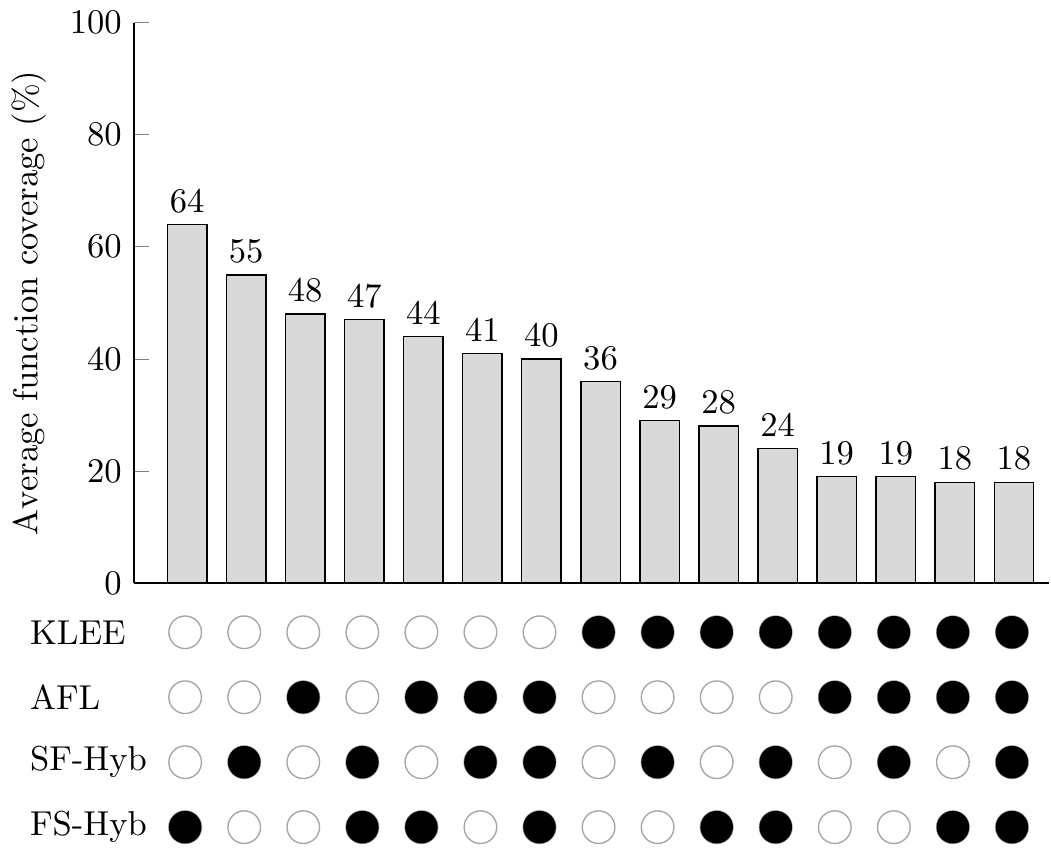}\vspace{-1em}
    \caption{Average function coverage by the four techniques and the intersection of covered functions between all techniques}
    \vspace{-1em}
    \label{fig:function-coverage}
\end{figure}

\paragraph{Function coverage}\label{sec:function-coverage}
We now evaluate the effectiveness of our method on real-world programs. \Cref{fig:function-coverage} shows the number of functions covered by fuzzing, symbolic execution and two described hybrid methods. It also shows the number of functions that were covered by all the indicated techniques. For example, the fourth bar in \cref{fig:function-coverage} shows that 47\% of the functions, on average, were covered by, \emph{both}, SF and FS hybrid. Similarly, 44\% of the functions, on average, were covered by, \emph{both}, AFL and FS hybrid, and so on.

Results for function coverage criteria are shown in \cref{tab:function-coverage}. 

\begin{table*}[h!]
    \centering
    \caption{Function coverage by four techniques}
    \label{tab:function-coverage}
    \begin{tabularx}{\textwidth}{@{}C{1}C{1}C{1}C{1}C{1}C{1}C{1}C{1}C{1}C{1}C{1}C{1.2}C{1.2}C{1.2}@{}}
     \toprule
      \multicolumn{3}{C{3}|}{} & \multicolumn{4}{C{4}|}{\hspace{1.5em}Coverage \newline(\%)} & \multicolumn{4}{C{4}|}{\hspace{2em}Time\newline(mins)} & \multicolumn{3}{C{3.6}}{\hspace{1em}Solver queries \newline(\#)} \\ \midrule
      \multicolumn{1}{C{1}|}{\vspace{0.4em}Prog.}        & \multicolumn{1}{C{1}|}{\vspace{0.4em}kLOC}    & \multicolumn{1}{C{1}|}{\vspace{-0.2em}\# functions}                & \vspace{0.4em}AFL & \vspace{0.4em}KLEE & \multicolumn{1}{C{1}}{\vspace{0.4em}FS} & \multicolumn{1}{C{1}|}{\vspace{0.4em}SF} & \multicolumn{1}{C{1}}{\vspace{0.4em}AFL} & \multicolumn{1}{C{1}}{\vspace{0.4em}KLEE} & \multicolumn{1}{C{1}}{\vspace{0.4em}FS} & \multicolumn{1}{C{1}|}{\vspace{0.4em}SF} & \vspace{0.4em}KLEE            & \vspace{0.4em}FS & \vspace{0.4em}SF \\ \midrule
        Bc                           & 3.5                    & 113                                  & 30  & 29   & 57    & 32                   & 300           & 300             &   113            & 120                       & 5227            & 2141             & 207 \\
        Bzip2                        & 3.3                    & 93                                   & 53  & 30   & 57  &   53                    & 300           & 300             &  100              & 120                       & 1430            & 167 & 167             \\
        Diff                         & 7.8                    & 197                                  & 39  & 23   & 58 &     23                   & 300           & 300             &  125             & 120                       & 9904            & 1918  &  65 \\
        Flex                         & 6.5                    & 217                                  & 68  & 18   & 70 &   68                     & 300           & 300             &   128            & 120                       & 151             & 27           &    14 \\
        Grep                         & 8.0                    & 192                                  & 13  & 19   & 39 &   28                    & 300            & 300             &  170             & 120                        & 41941           & 4229      &   29984    \\
        Jq                           & 8.9                    & 406                                  & 74  & 40   & 76 &    58                  & 300           & 300             &   101            & 120                       & 3505            & 1921     &    6    \\
        Lz4                          & 4.7                    & 94                                   & 24  & 69   & 73 &   57                      & 300           & 300             &  126             & 120                       & 3556                & 1969  &    97       \\
        Sed                          & 3.2                    & 125                                  & 42  & 73   & 60 &  86                         & 300           & 300             &  110             & 120                       & 23149           & 1684   &   21021       \\
        Zopfli                       & 1.9                    & 99                                   & 92  & 18   & 92 &   89                       & 300           & 300             &   65            & 120                       & 79296           & 1507   &  46         \\ \midrule
        \textbf{Avg.} & \textbf{5.3} & \textbf{170} & \textbf{48} & \textbf{36} & \textbf{64} & \textbf{55} & \textbf{300} & \textbf{300} & \textbf{115} & \textbf{120} & \textbf{18684} & \textbf{1729} & \textbf{5734} \\
        \bottomrule
    \end{tabularx}
    \vspace{-0.8em}
\end{table*}

We see in the table that the number of functions covered by our proposed hybrid methods was mostly higher than both, fuzzing and symbolic execution when used just by themselves. This is especially remarkable considering that the total time taken by both variants of hybrid technique is \emph{less than half} than that taken by fuzzing or symbolic execution alone. 
This can be trivially explained for fuzzing in the same way as the explanation given in \cref{sec:motivation}. 

The intuition for a higher function coverage than pure symbolic execution is as follows. At every branching condition in a program, symbolic execution adds it (and its negation) to the current path-condition and solves it (using SMT solvers) to determine whether the condition is satisfiable. Many times, such as for frontier nodes, these path-conditions become too complex for the SMT solver to solve in a reasonably short amount of time and, hence, the symbolic execution engine does not have enough time to explore all the nodes below frontier nodes in a call-graph. 

Fuzzing tools, such as AFL, on the other hand, generate many input mutations in a short amount of time so that, even though many branches remain unexplored, all functions on any program path, starting from \texttt{main} to a leaf node, are covered. If the mutation strategy is fast enough, as is the case with AFL, it becomes more likely that the random mutations generate enough diversity in the inputs to cover more unseen functions, compared to a strategy which blocks execution until the satisfiability of all branching conditions along the way, is proven. 

Despite the overall higher coverage achieved by our hybrid methods than symbolic execution or fuzzing, we observed for a few programs that the number of functions that were \emph{only} covered by hybrid method and not by the other two was rather small. Even though we do not accept this as a failure on the part of hybrid methods (because symbolic execution or fuzzing, by themselves, are not able to find as many functions as our technique does), we discuss in \cref{sec:discussion} the limitations of our fuzzing and symbolic execution engines that lead to this gap in coverage by the hybrid technique. 

\paragraph{Coverage depth}\label{sec:coverage-depth}
In the previous subsection, we discussed reasons for higher function coverage with hybrid fuzzing and targeted symbolic execution. 
However, the number of functions covered is not enough to demonstrate the benefits of a hybrid approach. Specifically, we show that, with our adaptive hybrid framework, it is possible to cover more functions in \emph{all depths} of a program's call-graph than either fuzzing or symbolic execution. We consider this to be an important metric because, in our experience, fuzzing misses a lot of vulnerabilities in ``lesser called'' functions of a program and symbolic execution misses them in deeper functions of a program that end up having low path coverage.

To demonstrate this effect, we performed an analysis on the covered and uncovered functions of all analyzed programs, w.r.t.\ the depth in the corresponding call-graph. Depth of a function in call-graph is defined with the usual definition of node depth in trees, i.e.\ the minimum number of edges from the node (function) to the root node (\texttt{main} function) of the tree. 
In this subsection, we aim to evaluate depth-wise coverage by our two hybrid approaches and compare it to those of fuzzing and symbolic execution.  

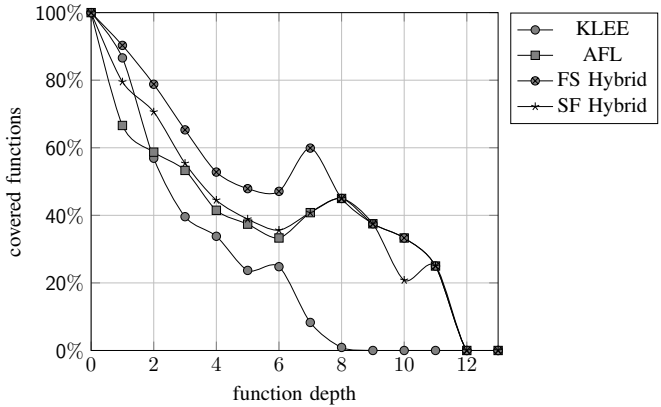
\begin{figure}[h]
    \vspace{-1.4em}
    \resizebox{\linewidth}{!}{%
        \providecommand{\myprogram}{avg}%
        \providecommand{\thispath}{figures}
        \begin{tikzpicture}
 \begin{axis}[
   title={\textbf{\phantom{Mg}\ifthenelse{\not{\equal{\myprogram}{avg}}}{\myprogram}{}}},
   xmin=0,
   ymin=0,
   ymax=100,
   xlabel={function depth},
   ylabel={covered functions},
   yticklabel={$\mathsf{\pgfmathprintnumber{\tick}\%}$},
   grid=major,
   enlargelimits=false,
   legend style ={ at={(1.03,1)}, 
       anchor=north west, draw=black, 
       fill=white,align=left},
    cycle list name=black white,
    smooth
  ]
  \addlegendentry{KLEE}
  \addplot table[y index=1] {\thispath/depth-cov-data/plot-\myprogram.dat};
  \addlegendentry{AFL}
  \addplot table[y index=2] {\thispath/depth-cov-data/plot-\myprogram.dat};
  \addlegendentry{FS Hybrid}
  \addplot table[y index=3] {\thispath/depth-cov-data/plot-\myprogram.dat};
  \addlegendentry{SF Hybrid}
  \addplot table[y index=4] {\thispath/depth-cov-data/plot-\myprogram.dat};
\end{axis}
\end{tikzpicture}%
    }%
    \vspace{-1em}
    \caption{Average (over all analyzed open-source programs) function coverage (\%) at all depths of call-graph(s)}
    \label{fig:coverage-depth}
\end{figure}

\Cref{fig:coverage-depth} shows the average function coverage at all depths of call-graph of all the analyzed programs. 
As per the figure, we can observe that 
\begin{enumerate*}
    \item Function coverage for symbolic execution (generally) degenerates with increasing call-graph depth. It is also the highest at low depths, and lowest at high depths. 
    \item Function coverage for fuzzing (generally) degenerates with increasing call-graph depth. However, at some point around medium depth, it increases and almost coincides with the hybrid technique's coverage. And,
    \item Function coverage with, both, FS and SF hybrid techniques also (generally) degenerates with call-graph depth. However, the steepness of coverage decline with these methods is lower than both, symbolic execution and fuzzing. 
\end{enumerate*}

From these observations, we may, firstly, reiterate our claim from sec.\ \ref{sec:function-coverage} that symbolic execution fails to cover many functions that lie at high depths of a call-graph because the execution is blocked until the satisfiability of each of the branching conditions on the way is determined by the SMT solver. 
Secondly, KLEE achieves much higher coverage than AFL at low depths of a program, due to the fuzzer's inability to ``guess'' the solutions to many branching conditions. This can also be explained by looking at the nature of our analyzed programs. Most of these programs are full of text parsing and/or sanitization in low depths of the program, before exposing the input buffer to the functionality. This is true for most programs that we analyzed, such as Grep, Flex, Sed and Diff. 
Thirdly, these results confirm our claim that a hybrid technique involving fast test-case generation (fuzzing) and deterministic path exploration (symbolic execution) should lead to better function coverage at all depths of the call-graph than either than either of the two involved techniques alone. 

Our results, however, raise a question about coverage achieved by FS hybrid at high depths of programs. It can be argued that by directly targeting functions not covered by fuzzing, targeted symbolic execution should perform \emph{at least} as well as symbolic execution alone. However, this was not always true for real-world programs that we analyzed, because of a technical drawback that we explain in \cref{sec:discussion}. 

\subsubsection{Efficiency}\label{sec:efficiency} 
\paragraph{Efficiency on artificial programs}\label{sec:efficiency-artificial-programs}\vspace{0.3em}
We first go back to the motivating examples of artificially generated programs in \cref{sec:motivation}. Recall that, as seen in \cref{tab:artificial-program-coverage}, fuzzing was unable to achieve acceptable function coverage for these programs. Symbolic execution, however, was able to cover all functions in all programs, but with an alarmingly high number of queries issued to the underlying SMT solvers. We also see in \cref{tab:artificial-program-coverage} that the total number of SMT solver queries issued by FS hybrid approach of Munch is consistently much lower than symbolic execution alone. Even though, qualitatively, these queries were sufficiently easy for the SMT solver to solve before timing out, these results indicate that our hybrid technique reduces the reliance on SMT solvers by a large amount -- thereby mitigating one of the two major drawbacks associated with symbolic execution, the other being path-explosion. 

Additionally, we found that, in many of the programs with high branching factors, \texttt{sonar-search} strategy made sure that many function nodes below frontier nodes (defined in \cref{sec:instrumentation}) in these programs were covered by KLEE in the same run as that targeting these frontier nodes. This trend indicates that with targeted symbolic execution, our hybrid technique may be able to find paths to interesting targets without getting ``stuck'' in shallow parts of a program due to path-explosion. 

A comparison with SF hybrid approach is excluded, because the first step of symbolic execution would cover all functions in a small amount of time, with the same number of queries as above.

\paragraph{SMT solver queries and analysis time}\label{sec:smt-queries-and-time} 
Now we consider the number of SMT queries generated by the symbolic execution engine and the total analysis time taken by the four compared techniques. Even though the number of generated queries may not directly correlate with analysis time, we consider this an important aspect because it points to the efficiency with which diverse paths (and, hence, functions) are covered in a complex program. We can see from \cref{tab:function-coverage} that the number of SMT queries dramatically drops with the FS hybrid, ranging from a drop by 45\% to 98\%. The number of queries was trivially lower with SF hybrid because the maximum time limit allowed here for symbolic execution itself was a fifth (1 hour) of symbolic execution alone. Therefore, due to an efficient implementation of \texttt{sonar-search} we were able to reduce the number of paths and, hence, the number of SMT queries generated by Munch.  

For a fair comparison of the techniques, however, we have to combine the number of SMT queries with total analysis time. With a timeout of \emph{2 minutes} per target function with \texttt{sonar-search}, FS hybrid was able to cover more functions than KLEE or AFL in less than half the total time, on average. SF hybrid, too, was able to cover more functions with only one hour of symbolic execution with KLEE and one hour of fuzzing (starting with more diverse seed inputs than AFL alone) than KLEE or AFL alone in less than half the total time, on average. 

\subsubsection{Summary}
With the above results, we were able to show that, when analyzing a program for a less amount of time,
\begin{enumerate*}
    \item Overall function coverage achieved by both, FS and SF hybrids, were always better than, or as high as, those achieved by fuzzing or symbolic execution alone. 
    \item Function coverage for deep functions achieved by our hybrid techniques beat symbolic execution or fuzzing alone, especially from the perspective of rate of drop in function coverage as we go deeper in the call-graph. 
    \item The number queries generated for SMT solver was reduced by 80\% on average.
    \item Using Munch framework, therefore, increases the chances that vulnerabilities may be spotted in more functions at all depths of the program rather than either only at low depths (as is the case with symbolic execution alone) or high depths but not for many branches (as is the case with fuzzing alone), using considerably fewer calls to the SMT solver and in much less time. 
\end{enumerate*}
\subsection{Discussion and Limitations}\label{sec:discussion}
We see from the evaluation presented above that a hybrid framework that combines fuzzing and symbolic execution leads to higher function coverage at a much lower cost than either of the two involved techniques employed alone. When we look at the coverage intersections in \cref{fig:function-coverage}, however, we can also observe that the hybrid technique does not always cover many functions that could not \emph{also} have been covered by AFL or KLEE alone. However, we must note that this limitation is not due to a conceptual flaw in the design, but due to limitations in KLEE. KLEE, with its default search strategy, only updates the paths to uncovered states \emph{once every 30 seconds}. Our search strategy, (\texttt{sonar-search}), on the other hand, always calculates precise distances, which leads to thousands of calculations that could not be cached across multiple runs of KLEE (one run for every target function) with \texttt{sonar-search}. However, for symbolic execution alone, KLEE is only run once on the whole program and, therefore, can cache SMT queries that have been solved in the past.
Without a caching mechanism working \emph{between} several runs, targeted symbolic execution spends most of the time analyzing initial parts of a program multiple times to calculate shortest paths to target functions. This limitation of KLEE, in fact, shows the hybrid method in an even better light when seen from the perspective of queries issued to the SMT solver (last three columns in \cref{tab:function-coverage}). Symbolic execution, by itself, was able to cache a lot of queries during execution which it did not have to issue to the SMT solver when encountered again (e.g., when popping a function from the stack). Even with this advantage over targeted search (having to issue fewer queries to SMT solver), KLEE takes more time than Munch to cover new functions, overall. 

AFL also comes with some design peculiarities that \emph{might} have contributed to low coverage in deep functions. The technical whitepaper of AFL \cite{afl} states that its execution schedule is optimized to not get stuck in those nodes in the control-flow graph that increase the path coverage by a small enough amount to not warrant further exploration. This avoidance aspect of AFL may be responsible for low coverage in high depths of call-graph during, both, SF and FS modes of operation. Another drawback of AFL that might affect Munch is its insistence on either \texttt{STDIN} or file based inputs, but not both. Because of this some functions in the program that may only be triggered by a certain combination of, both, command line arguments and file inputs are not covered with either hybrid modes. 

Therefore, we have shown in our results that hybrid approaches, as implemented in Munch,  involving symbolic execution and fuzzing may be, in the context of the analyzed programs, more effective and efficient in terms of deeper function coverage than fuzzing or symbolic execution alone.

\section{Related Work}\label{sec:background} 
\paragraph{Fuzzing}\vspace{0.3em}
Fuzzing was first used as a variant of randomized testing for the reliability of UNIX tools by \citeauthor{miller1990empirical} \cite{miller1990empirical}. Since then, many fuzzing (or blackbox fuzzing) implementations, such as \cite{eddington2011peach,aitel2002introduction}, have been developed for different target-systems and applications. One of the most popular fuzzers in academia and industry currently is AFL \cite{afl}, which uses genetic algorithms and instrumentation for an improving path coverage and vulnerability discovery.
 
\paragraph{Symbolic execution}\vspace{0.3em}
Symbolic execution, introduced by \citeauthor{king1976symbolic} \cite{king1976symbolic}, deterministically analyzes a program by representing its paths as constraint systems that are solved (using constraint solvers) to generate test cases executing those paths. Concolic execution \cite{cadar2005execution}, whitebox fuzzing \cite{godefroid2008automated} and bounded model checkers (BMC) \cite{biere2003bounded} are some practical approaches to symbolic execution that have been proposed in the past. 
Due to a whitebox view of the program-under-test, symbolic execution is, theoretically, able to cover more diverse branches than fuzzing, but suffers from two particular problems -- path-explosion \cite{cadar2013symbolic} and inefficient constraint solving \cite{anand2008demand,pretschner2001classical}. To overcome path-explosion, many approaches have been proposed such as \cite{kuznetsov2012efficient,sen2015multise,godefroid2007compositional}, out of which compositional analysis is one. Compositional symbolic execution, which is a building block of our technique, breaks down a large program into smaller components (e.g.\ functions) such that these components can be analyzed in isolation from one another, thereby reducing the number of paths to be explored, and their results are composed together. Various evaluations of compositional symbolic execution approach \cite{christakis2015ic,ognawala2016macke,qiu2015memoized} have shown that it is more effective at increasing coverage and finding low-level vulnerabilities than forward symbolic execution. 
 
\paragraph{Hybrid methods}\vspace{0.3em}
The main distinction between Munch and most other hybrid techniques that have been proposed over the past decade is that our framework can be used in \emph{two} different modes of operation based on the availability of diverse test-cases. An early hybrid of symbolic execution and random testing was introduced by \citeauthor{majumdar2007hybrid} in 2007 \cite{majumdar2007hybrid} where the authors proposed applying concolic execution after random testing reaches a saturation point. However, robust implementations of fuzzing (successor of random testing) were only introduced in later years, so as to make the above technique more efficient. In \cite{fangquan2015binary} a hybrid method is introduced which employs symbolic execution to increase the ``breadth'' of coverage in binaries when fuzzing does not increase the basic-block coverage. Our FS hybrid improves this by directing execution towards the uncovered basic-blocks or functions and, therefore, avoiding those blocks which had already been covered with fuzzing. In \cite{pak2012hybrid}, \citeauthor{pak2012hybrid} develops and evaluates an equivalent of SF hybrid strategy, but not FS. \citeauthor{wang2010taintscope} \cite{wang2010taintscope} present a hybrid approach applicable for programs dealing with input-dependant checksum operations. Our framework, in principle, works for any C program, including the ones containing checksum operations. \citeauthor{bottinger2016deepfuzz} \cite{bottinger2016deepfuzz} propose a hybrid method that is able to find vulnerabilities deep inside a program. However, the difference between \cite{bottinger2016deepfuzz} and our paper is that the symbolic execution search strategy is guided in \cite{bottinger2016deepfuzz} by assigning likelihoods to paths that may execute yet undiscovered (possible) vulnerabilities, and we use instrumentation and function coverage for the same. In \cite{pham2016model} the authors introduce a whitebox fuzzing strategy, called \emph{Model-based Whitebox Fuzzing (MoWF)}. MoWF uses model-based file description to overcome the shallow parts of any program that guard their specific (deep) functionality and then uses symbolic execution. Like target-function guided symbolic execution, input models can also help in reducing the constraint space. None of the above works are accompanied by open-source implementations, so a comparison w.r.t.\ our evaluation metrics proved difficult. 
 
In Driller \cite{stephens2016driller}, which is the closest to our framework, \citeauthor{stephens2016driller} discover the same patterns about coverage-in-depth as our work, and, hence, are able to find many more vulnerabilities than na\"{i}ve symbolic execution and fuzzing tools. However, their symbolic execution exploration strategy (after initial fuzzing) is guided by a taint analysis on the inputs generated by the fuzzer, instead of a target-based directed strategy. Because of relying on the inputs generated by the fuzzer, instead of pre-calculating a call-graph and computing a difference set to determine targets of interest (uncovered functions or basic blocks), Driller does not guarantee higher function coverage, in general. We tried to apply Driller to our evaluation set but were unable to do so because its open-source code base is tailor-made for the authors' distributed system set-up, and not ready to use out-of-the-box. Munch, on the other hand, is ready to be installed and used on any Linux-based system. 
\section{Conclusion}\label{sec:conclusion}
In this paper, we have developed and evaluated Munch, a hybrid framework for function coverage, based on fuzzing and symbolic execution. This adaptive framework works in two modes -- fuzzing followed by targeted symbolic execution and symbolic execution to generate diverse seed inputs for fuzzing. Using our implementation of targeted search in symbolic execution, the number of queries to the SMT solver, and, therefore, the time taken for symbolic execution, is reduced by only focussing on those paths that may lead to uncovered functions.

We demonstrated the efficiency of Munch on 12 artificially generated programs. We also evaluated it on 9 widely used open-source programs for effectiveness and efficiency and compared them to pure fuzzing and symbolic execution. We found that, as shown in \cref{sec:results}, our hybrid approach successfully achieves higher function coverage at all depths of a program's call-graph than either fuzzing or symbolic execution alone, at less than half of the cost (time and SMT queries). Looking at these results, we can posit that, with higher function coverage in highly compositional programs, our hybrid technique is more likely to find vulnerabilities that fuzzing or symbolic execution may miss when used by themselves. 

\bibliographystyle{plainnat}
\bibliography{literature}
\end{document}